\documentclass[aps,prl,reprint,a4paper,superscriptaddress]{revtex4-1}

\usepackage{graphicx}

\usepackage{epstopdf}
\usepackage{amsmath}
\usepackage{siunitx}
\usepackage{color}
\usepackage{natbib}
\usepackage{booktabs}
\usepackage{cleveref}
\usepackage[colorlinks=true]{hyperref}

\definecolor{citecol}{rgb}{0.0, 0.6, 0.2}
\definecolor{filecol}{rgb}{0.2, 0.8, 0.8}
\definecolor{linkcol}{rgb}{0.8, 0.2, 0.1}
\definecolor{urlcol}{rgb}{0.9, 0.1, 0.6}
\definecolor{revcol}{rgb}{0.4, 0.1, 0.0}
\hypersetup{colorlinks,citecolor=citecol,filecolor=filecol,linkcolor=linkcol,urlcolor=urlcol}

\newcommand{\ket}[1] {\left|{#1}\right>}

\newcommand{\ssection}[1]{} %temporary headings off
\newcommand{\rrevised}[1]{#1} %revision colour off

\begin{document}

%%%%%%%%%%%%%%%%%%%%%%%%%%%%%%%%%%%%%%%%%%%%%%%%%%%%%%%%%%%%%%%%%%%%%

\title{Addressable electron spin resonance using donors and donor molecules in silicon}

%\author{Sam \& Co}%

\author{Samuel J Hile}
 \email{samhile@gmail.com}
\author{Lukas Fricke}%
\author{Matthew G House}%
\author{Eldad Peretz}%
  \affiliation{%
Centre for Quantum Computation and Communication Technology ($CQC^2T$),\\ School of Physics, University of New South Wales, Sydney 2052, Australia
}
\author{Chen Chin Yi}%
\author{Yu Wang}%
  \affiliation{%
Network for Computational Nanotechnology, Purdue University, West Lafayette, IN, USA
}
\author{Matthew Broome}%
\author{Samuel K Gorman}%
\author{Joris  G Keizer}%
  \affiliation{%
Centre for Quantum Computation and Communication Technology ($CQC^2T$),\\ School of Physics, University of New South Wales, Sydney 2052, Australia
}
\author{Rajib Rahman}%
  \affiliation{%
Network for Computational Nanotechnology, Purdue University, West Lafayette, IN, USA
}
\author{Michelle Y Simmons}%
 \email{michelle.simmons@unsw.edu.au}
 \affiliation{%
Centre for Quantum Computation and Communication Technology ($CQC^2T$),\\ School of Physics, University of New South Wales, Sydney 2052, Australia
}%

%%%%%%%%%%%%%%%%%%%%%%%%%%%%%%%%%%%%%%%%%%%%%%%%%%%%%%%%%%%%%%%%%%%%%
%% Some journals require a list of abbreviations or keywords to be
%% supplied. These are here if needed.
%%%%%%%%%%%%%%%%%%%%%%%%%%%%%%%%%%%%%%%%%%%%%%%%%%%%%%%%%%%%%%%%%%%%%
\keywords{Electron Spin Resonance, Hyperfine Coupling, Single Donor, Silicon, Quantum Computing, Addressable Qubits}

%%%%%%%%%%%%%%%%%%%%%%%%%%%%%%%%%%%%%%%%%%%%%%%%%%%%%%%%%%%%%%%%%%%%%
%% Internal Comments - remove for publication
%%%%%%%%%%%%%%%%%%%%%%%%%%%%%%%%%%%%%%%%%%%%%%%%%%%%%%%%%%%%%%%%%%%%%
%{Revised changes in \rrevised{brown}}
%%%%%%%%%%%%%

%%%%%%%%%%%%%%%%%%%%%%%%%%%%%%%%%%%%%%%%%%%%%%%%%%%%%%%%%%%%%%%%%%%%%
%% Abstract - avoid maths here
%%%%%%%%%%%%%%%%%%%%%%%%%%%%%%%%%%%%%%%%%%%%%%%%%%%%%%%%%%%%%%%%%%%%%

\begin{abstract}

Phosphorus donor impurities in silicon are a promising candidate for solid-state quantum computing due to their exceptionally long coherence times and high fidelities. However, individual addressability of exchange coupled donor qubits with separations $\sim 15\si{nm}$ is challenging. Here we show that by using atomic-precision lithography we can place a single P donor next to a 2P molecule $16 \pm 1 \si{nm}$ apart and use their distinctive hyperfine coupling strengths to address qubits at vastly different resonance frequencies. In particular the single donor yields two hyperfine peaks separated by $97 \pm 2.5 \si{MHz}$, in contrast to the donor molecule which exhibits three peaks separated by $262 \pm 10 \si{MHz}$. Atomistic tight-binding simulations confirm the large hyperfine interaction strength in the 2P molecule with an inter-donor separation of $\sim 0.7 \si{nm}$, consistent with lithographic STM images of the 2P site during device fabrication. We discuss the viability of using donor molecules for built-in addressability of electron spin qubits in silicon.

\end{abstract}

\maketitle


\section{Introduction}

%Scalability remains a principal challenge in the development of practical quantum information technologies.
Phosphorus donor atoms in silicon are very attractive as the basis of a solid-state quantum computer, because they combine the long-lived quantum memory of a nuclear spin with the rapid control and strong interactions possible with an electron spin \cite{kane1998silicon}.
With weak coupling to their environment, \rrevised{phosphorus donors in isotopically purified $^{28}$Si have demonstrated minutes-long nuclear \cite{steger2012quantum} and millisecond-long electron \cite{tyryshkin2012electron} spin coherence times.}
Additionally, high fidelity single-qubit quantum gate operations using resonant magnetic fields \cite{muhonen2015quantifying} and high fidelity state readout \cite{watson2017atomically,pla2013high} have recently been demonstrated for single P donor qubits.

Due to the strong Coulomb potential well, donors provide a means of producing uniform electron spin qubits with reproducible tightly confined wavefunctions, and thus a non-degenerate, low-lying valley and orbital ground-state.
To achieve an accurate entangling two-qubit quantum gate, via the exchange \cite{koiller2001exchange,song2016statistical} or dipole \cite{de2004silicon} interaction, donors must be placed with high precision and on the order of tens of nanometres apart. We achieve this fine positional control through scanning tunnelling microscope (STM) hydrogen resist lithography \cite{schofield2003atomically} which has an effective resolution less than the silicon lattice constant. One of the challenges therefore, is to individually address nominally identical qubits when they are very close together.
\rrevised{In gate-defined silicon quantum dot single-spin qubits, typically placed $\sim100\si{nm}$ apart, the nesseccary addressability in electron spin resonance has been achieved via} a slanting Zeeman field generated by a surface micromagnet \cite{noiri2016coherent,noiri2016coherent}, or via a Stark shift in the electron spin resonances \cite{veldhorst2014addressable}.
Similar tunability of qubit resonance frequencies has been observed for a single donor qubit \cite{laucht2015electrically}. However, in moving to multi-qubit systems, with $\sim 15 \si{nm}$ inter-donor separations, a higher B- or E-field gradient will be needed to avoid overlapping qubit resonance frequencies \cite{buch2013spin}.
%, even with power-limited linewidths in $^{28}$Si substrates.

In this paper we demonstrate the successful implementation of an alternate strategy for addressing donor-bound electron spin qubits. Here we use the differentiated hyperfine coupling of an electron confined by the potential well formed by a single P donor and by a pair of donors in a donor molecule \cite{buch2013spin,wang2016characterizing}. We present electron spin resonance (ESR) measurements of the hyperfine spectrum of both a single donor (1P) and a donor molecule (2P) within a single double quantum dot device.

The spin states of each individual quantum dot may be described by the generalised Hamiltonian containing an electron Zeeman term and nuclear Zeeman terms as well as hyperfine interaction terms for each donor nucleus, $i$.
\begin{gather*}
 H = g_e \mu_B B_0 S_z
 + \sum_i{g_n \mu_N B_0 {I_i}_z}
 + \sum_i{\vec{S} \cdot \left( A_i\mathbf{I} + \mathbf{D_i} \right) \cdot \vec{I_i}}
 \end{gather*}
Here $g_e$ ($g_n$) is the electron (nuclear) g-factor, $\mu_B$ ($\mu_N$) the Bohr (nuclear) magneton, and $S$ ($I$) the electronic (nuclear) spin angular momentum.
The static magnetic field $B_0$ is oriented parallel to the patterned surface, and aligned with the $\left[110\right]$ crystal axis, and \rrevised{the hyperfine coupling tensor for each nuclei $i$ is decomposed into a scalar Fermi contact interaction part $A_i$ and an anisotropic dipolar component $\mathbf{D_i}$, which is often treated as negligible.}

\begin{figure*}[hbtp]
\includegraphics[width=0.98\textwidth]{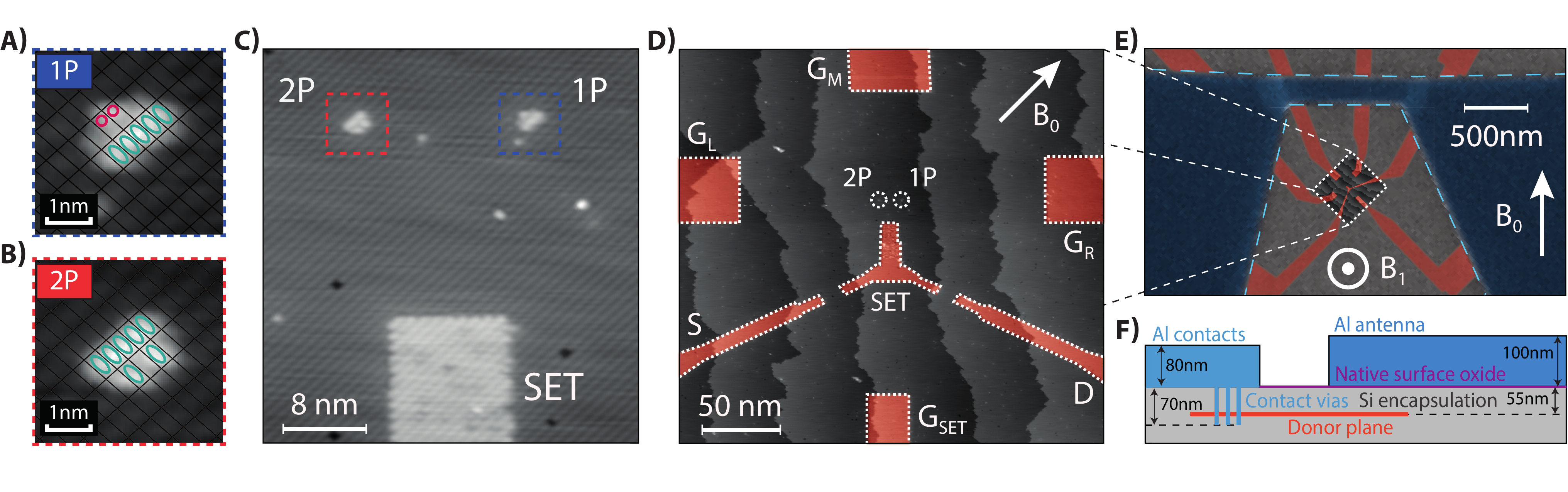}
\caption{\label{fig:one} \textbf{Alignment of a surface ESR antenna to the buried atomic precision double quantum dot.}
(A) Atomic resolution STM image of the single donor (1P) quantum dot and (B) donor molecule quantum dot (2P). The $2\times 1$ surface reconstruction of dimers is indicated by a rectangular black grid, with reactive exposed dimers highlighted in green, and single unreactive exposed Si atoms marked in pink.
(C) High resolution STM image of the two quantum dots and the readout SET, and (D) larger-scale image showing the full device structure. False-coloured red areas are phosphorus doped to form the sensor SET, source and drain leads, and electrostatic gates.
(E) False-colour composite SEM and STM image showing the buried donor structures (red) relative to the aluminium surface antenna (blue) which generates an oscillating $B_1$ field out of the plane as indicated. The direction of the static $B_0$ field produced by a superconducting magnet is also shown.
(F) Vertical cross-section, showing the thickness (not to scale) and relative position of the silicon, phosphorus, oxide and aluminium layers. The Al contacts are not seen in (A-E), as they are positioned $>1\si{\mu m}$ away from the antenna.
}
\end{figure*}

\section{Results}

\subsection{Alignment of ESR antenna}

The device presented is a planar donor-defined nanostructure, in which 2D regions of a silicon crystal are heavily phosphorus-doped beyond the metal-insulator transition by STM hydrogen resist lithography, to produce gates, reservoirs and also localised charge islands \cite{schofield2003atomically}.
Figs. 1A,B show the lithographic mask at the two quantum dot sites, relative to the $2\times1$ dimer reconstruction on the silicon surface during fabrication.
To incorporate a single P atom, three adjacent exposed dimers are required. This permits the phosphine gas ($PH_3$) molecule to fully dissociate at the surface \citep{wilson2006thermal}. From the images (both showing 5 consecutive exposed dimers), we expect at least one P atom in each quantum dot, with a possibility for two at the left site due to the presence of multiple additional exposed dimers nearby.
The actual number of donors incorporated at each site was subsequently verified by measuring each quantum dot's charging energy. Here we determined the single electron addition energies to be $E_{2P} \sim 65$meV and $ E_{1P} \sim 43$meV, consistent with NEMO3d tight-binding simulations reflecting 2P and 1P for the left and right quantum dot respectively \cite{weber2014spin}. These assignments are further confirmed by the ESR results following.
The two sites are separated by $16\pm1$nm, and are both tunnel-coupled (at a distance of 19nm) to a larger charge-sensing single electron transistor (SET) as shown in Fig. 1C, for energy selective spin readout \cite{elzerman2004single}. The SET additionally functions as an electron reservoir for the two donor sites.
The full layout including electrostatic gates is shown in Fig. 1D, where red regions represent metallic conductive structures of delta-doped epitaxial silicon, with a carrier density of $n=\num{2.5e14} \si{cm^{-2}}$ \cite{mckibbin2010investigating}. The conducting phosphorus structures are buried below a 55nm thick encapsulation layer, and contacted by etching vias and depositing aluminium surface contacts.

\ssection{Antenna Integration}

Following initial characterisation of the device including independent spin readout and spin correlation measurements \cite{broome2017}, a broadband microwave antenna was post-fabricated on the chip.
This is a remarkable feature of donor-based all-epitaxial devices. Since the dopant layer is protected by the crystalline silicon environment, which is conductive ($\rho \sim 10 \si{\ohm cm}$) at room temperature due to background doping, electrostatic discharge is unlikely. Hence these devices can be measured at cryogenic temperatures (where background dopants freeze out) multiple times, and be re-processed to add additional surface gates, waveguides or antennas, before being measured again.

The antenna geometry is impedance matched \cite{dehollain2012nanoscale} to minimise radiative and reflective loss of microwave power, whilst maximising the oscillating magnetic field, $B_1$. The post-fabrication process requires additional electron beam lithography, achieved with positional uncertainty of $<200$nm relative to the buried atomic scale device, by reference to pre-etched alignment markers \cite{ruess2005use}. Physical vapour deposition of 100nm of aluminium onto the naturally oxidised silicon surface produces an antenna capable of withstanding up to 2V DC bias relative to the buried phosphorus layer with minimal current leakage ($R>100\si{G\Omega}$).

The inner region of the completed antenna is seen in Fig. 1E (coloured blue), positioned with the donors inside the loop of the antenna where the simulated ratio of oscillating magnetic field, $B_1$, to in-plane oscillating electric fields is maximised (see Supplementary material I). A vertical cross-section of the device structure is shown in Fig. 1F.

\subsection{Addressable resonance spectra}

We operate in the high magnetic field regime ($B>1.2$T) such that $g_e \mu_B B_0 > A > g_n \mu_N B_0$, where the eigenstates are to first order separable into electron and nuclear subspaces, and we perform our measurements at $\sim 50\si{mK}$ in a dilution fridge.
The spin resonance experiment proceeds by applying voltages proportionally to the left and right gates, in order to detune the donor potential relative to a fixed SET Fermi energy (see Supplementary material II). Since the electron spin relaxation time is much longer than the characteristic tunneling time between SET and donor, we initialise the state by ionising the donor and deterministically loading an electron in the spin down state $\ket{\downarrow}$. The donor-bound electron is in Coulomb blockade whilst a microwave pulse is applied. This is followed by a single-shot readout sequence in which spin dependent tunneling \cite{elzerman2004single,buch2013spin} converts the projected electron spin state to a charge state, observable via the SET current signal.

The sequence used for the 2P molecule is equivalent, but instead of conditionally ionising the molecule by removing the single electron, it utilises a spin-dependent transition into the 2 electron spin singlet state \cite{watson2015high} for readout. This technique has the additional benefit of faster tunnelling rates, which permits faster operation relative to the single donor (1P).
Our microwave ESR pulses are applied with a nominal power at the signal generator of $+5\si{dBm}$ (we estimate $\sim 60\si{dB}$ attenuation at the device) for $150\si{\mu s}$, and modulated with a linear frequency chirp of $\pm 20 \si{MHz}$.
This adiabatic passage pulse \cite{garwood2001return} inverts the electron spin eigenstates irrespective of the exact pulse duration or precise instantaneous resonance frequency, enhancing our spin resonance signal \cite{laucht2014high}.

\begin{figure}[hbtp]
\includegraphics[width=0.48\textwidth]{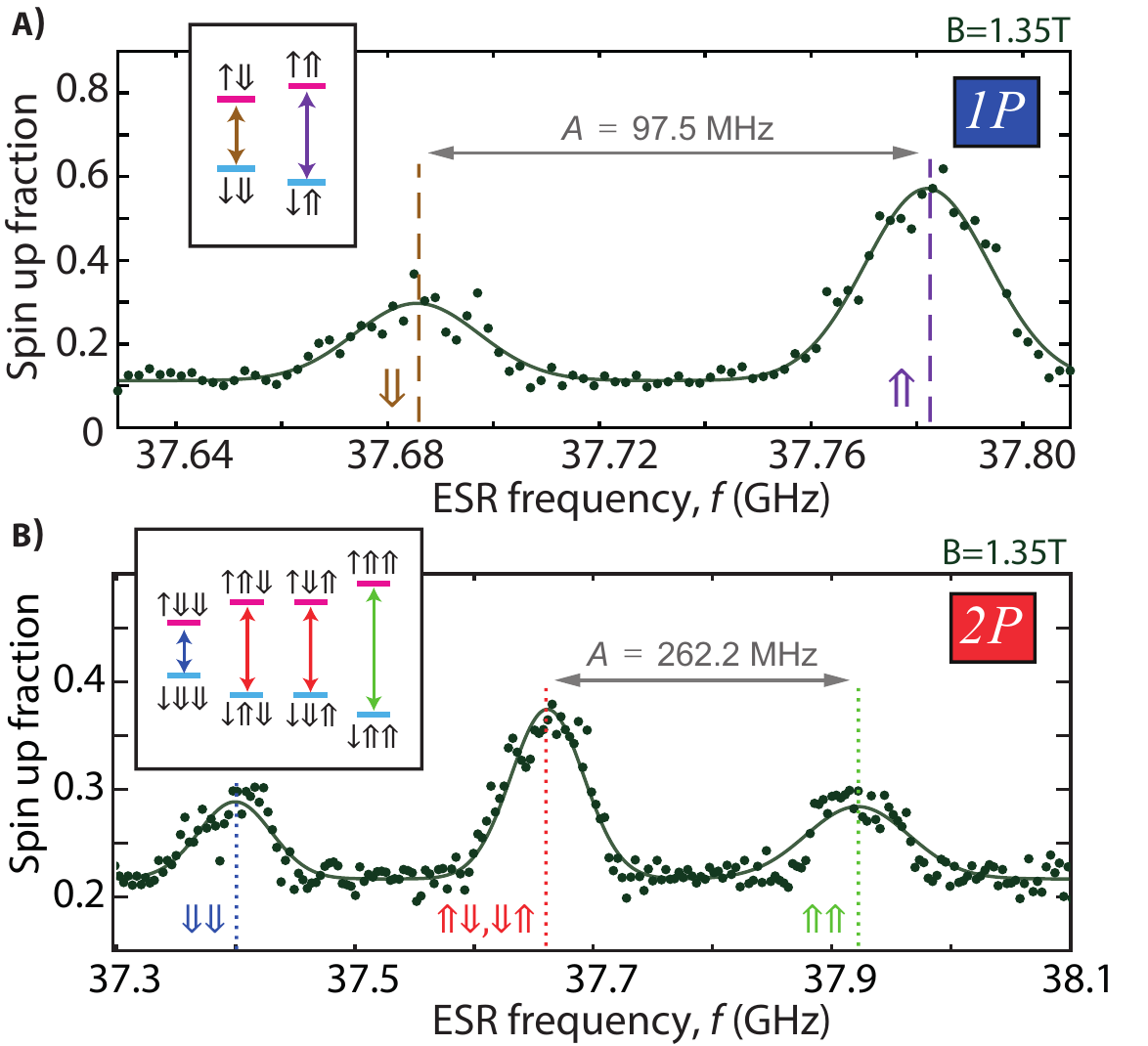}
\caption{\label{fig:two} \textbf{Electron spin resonance spectra for a single P donor and a 2P molecule}
(A) Measured ESR spectrum for the 1P bound electron, and
(B) 2P donor molecule at $B_0=1.35T$ using a $\pm 20$MHz adiabatic passage frequency chirp.
Insets indicate the nuclear (double arrow) and electron (single arrow) spin eigenstates and the ESR transitions between them, each corresponding to an observed resonance peak. The measured hyperfine energies $A_{1P}, A_{2P}$ are indicated for each donor quantum dot.
}
\end{figure}

\ssection{Spin Resonance Spectra}

The spin resonance spectrum of the single donor (1P) is shown in Fig. 2A for $B_0 = 1.35\si{T}$. This data shows the fraction of $\ket{\uparrow}$ outcomes, $p_{\uparrow}$, over 640 single-shot measurements at each frequency, $f$.
We observe two peaks, corresponding to the resonance conditions for driving transitions between electron $\ket{\downarrow}$ and $ \ket{\uparrow}$ states when the single nuclear spin state is either $\ket{\Downarrow}$ (left peak) or $\ket{\Uparrow}$ (right peak). The transition frequencies here are:
$f_\Downarrow = g_e \mu_B B_0 - \frac{A_{1P}}{2}$ and $f_\Uparrow = g_e \mu_B B_0 + \frac{A_{1P}}{2}$, separated by the single donor hyperfine coupling strength $A_{1P}$.
The solid curve in Fig. 2A is a fit to the sum of two Gaussian peaks sharing a common full width at half maximum $\Delta f_\text{FWHM} = 27.8 \pm 2\si{MHz}$
and with amplitudes $p_\Downarrow = 0.18$ and $p_\Uparrow = 0.46$.

\rrevised{The hyperfine coupling strength $A_{1P} =f_\Uparrow-f_\Downarrow= 96.5\pm2.5 \si{MHz}$ for the single P atom is comparable to other values in the literature for ion-implanted P donor devices ranging from $96.9$ to $116.6\si{MHz}$\cite{muhonen2015quantifying,laucht2015electrically}.  The difference between this and the value of the hyperfine reported for bulk ensembles of P donors $\sim 117.5\si{MHz}$\cite{feher1959electron1} can be attributed to a Stark shift, since within our device we have an electric field at the 1P site of $E \sim 4.5\si{MV/m}$ (see Supplementary Material III), perturbing the wavefunction. At this E-field, the reduced the electron density over the nucleus leads to a quadratic Stark effect in P donors \cite{pica2014hyperfine,rahman2007high} and thus $A_{1P}$ is reduced by a factor $\num{2.5e-3}\si{(MV/m)^{-2}}$, consistent with our measurement. }

We note that in Fig. 2A the $\ket{\Uparrow}$ peak has more than twice the amplitude of the $\ket{\Downarrow}$ resonance, indicating that there is some polarisation of the nuclear spin, with the $\ket{\Uparrow}$ state more likely to be occupied than $\ket{\Downarrow}$.
This polarisation reflects nuclear spin dynamics likely arising from an inelastic electron-nuclear flip-flop process, pumped by spin resonant excitation at the $f_{\Downarrow}$ frequency \cite{pines1957nuclear,mccamey2009fast}.
Here an electron spin `flips' from $\ket{\uparrow}$ to $\ket{\downarrow}$, and the nuclear spin simultaneously `flops' from $\ket{\Downarrow}$ to $\ket{\Uparrow}$. The total spin is thus conserved, and energy conservation is satisfied by the emission of a phonon. Since the energy difference between the states is larger than the thermal energy $g_e \mu_B B_0 > k_B T$, the reverse transition involving absorption of a phonon is suppressed. Any alternate cross-process involving the $\ket{\downarrow\Downarrow} $ and $\ket{\uparrow\Uparrow}$ states would require a change in total spin of $\pm1$ and so is forbidden by spin conservation.\rrevised{ Interestingly, we infer from fluctuations in our recorded spin-up signal over time  (see Supplementary material IV for further analysis on the nuclear dynamics) that the timescale for the flip-flop process may be as short as $50\si{s}$, orders of magnitude faster than expected for P donors at our magnetic field and temperature\cite{pines1957nuclear,mccamey2009fast}. We may attribute this increased rate to an enhancement in the electron-phonon interaction due to the non-trivial valley structure of the electric field-perturbed wavefunction inside our nanostructure\cite{boross2016valley}.}
Partial repopulation of the $\ket{\Downarrow}$ state can be explained by an `ionisation shock' \cite{pla2013high}, where mis-alignment of the nuclear spin eigenstates in the neutral and ionised donor charge states provides a small non-zero probability of flipping the nuclear spin on each electron ionisation event. 

In Fig. 2B we present the resonance spectrum measured for the electron bound to the 2P molecule. This data is based on 2000 single shots (compared to 640 for the 1P single donor due to the faster tunnel rate between the SET and 2P molecule), with a microwave pulse power of $+8\si{dBm}$ and the same chirp parameters as above.
Here we observe three resonant frequencies. The solid curve is a fit to three Gaussian peaks of width $\Delta f_\text{FWHM} = 72\pm5\si{MHz}$
and amplitudes: $p_{\Downarrow\Downarrow} = 0.072,  p_{\Uparrow\Downarrow/\Downarrow\Uparrow} =  0.158$ and $p_{\Uparrow\Uparrow} = 0.067$.
The three peaks reflect the transition frequencies, shown in the inset to Fig. 2B, separated by $A_a$ and $A_b$, the contact hyperfine interaction coefficients representing the electron wavefunction density at the location of the two donor sites (labelled $a$ and $b$) of the 2P molecule. At zero electric field the $2P$ electron wavefunction is symmetric and the hyperfine interaction at the two donor sites is expected to be equal $A_a = A_b$, producing two degenerate transition frequencies $f_{\Downarrow\Uparrow} = f_{\Uparrow\Downarrow}$.
However, at the operating point of the 2P molecule in our device, we calculate an electric field of around $4.3\si{MV/m}$ (see Supplementary material III, and VII), which serves to break this degeneracy $A_a \neq A_b$. Since we resolve only one central peak in the resonance spectrum, the hyperfine asymmetry $|A_a - A_b|$ must be less than the width of the observed peak $\sim 72\si{MHz}$.
The average peak separation, representing the donor molecule hyperfine interaction energy $A_{2P}=(A_{a}+A_{b})/2 = 261\pm10\si{MHz}$ is more than twice the single donor value, consistent with the anticipated range for a pair of donors with small ($<1\si{nm}$) spatial separation \cite{wang2016characterizing}.
\rrevised{We can calculate the Stark shift expected for the 2P molecule for the same electric fields (see Supplementary Material VII). The value ranges from $<10\si{kHz}$ (with the electric field perpendicular to the molecular axis of the 2P molecule) to $6\si{MHz}$ (with the electric field parallel to the molecular axis of the 2P molecule). }

Interestingly, we note that the asymmetry in the peak amplitudes seen in the 1P case is absent in the 2P molecule's ESR spectrum. In Fig. 2B we see an equal probability for each of the 4 nuclear spin state resonances. The nuclear $\ket{\Downarrow\Uparrow}$ and $\ket{\Uparrow\Downarrow}$ states are approximately degenerate and indistinguishable, producing a peak with approximately twice the amplitude of the $\ket{\Downarrow\Downarrow}$ and $\ket{\Uparrow\Uparrow}$ resonances.

To understand the nuclear spin dynamics in the 2P molecule, we consider the full tensor form of the hyperfine interaction $\sum_{i=a,b}{ A_i \mathbf{I} + \mathbf{D_i}}$, which for each of the donors consists of the Fermi-contact hyperfine scalar $A$, proportional to the electron wavefunction density at the position of a donor atom and the traceless dipole-dipole interaction tensor $\mathbf{D}$ (see Supplementary material V). Since donor atoms have a strong Coulomb confinement, the electron wavefunction is highly concentrated over the donor nuclei, and $A$ dominates by several orders of magnitude over $\mathbf{D}$ for a single donor, even with significant perturbation by an electric field \cite{pica2014hyperfine,park2009mapping}.
The dipolar tensor is expected to be more anisotropic in the case of the molecular 2P wavefunction \cite{saraiva2015theory}, since it is inherently non-spherical.
Indeed we find that there is no correlation in the nuclear spin state between successive electron readout events, confirming a significant enhancement of the `ionisation shock' mechanism relative to that observed for the single donor. In short, the nuclear spin state in the 2P molecule is randomised faster than it is polarised by any inelastic relaxation.
%A potential strategy to mitigate the ionisation shock effect may be to align the external magnetic field with the symmetry axis of the molecular wavefunction.

\rrevised{Next we turn to consider the disparity in resonance linewidths in the 1P and 2P cases, important as this reflects the coherence properties of the bound electron spin. The 1P resonance peaks have a linewidth of $27.8\si{MHz}$ 8MHz as a result of being artificially broadened by the $20\si{MHz}$ linear frequency chirp we apply to our microwave pulse to adiabatically invert the spin state. The spin dephasing time for a single P donor in natural silicon was recently measured as 55ns \cite{pla2012single} corresponding to a natural linewidth (FWHM) around $10\si{MHz}$. This is limited by random fluctuations in the local magnetic field due to presence of $^{29}\si{Si}$ nuclear spins. We use the adiabatic pulse strategy [24,25] to combat the fluctuations and selected the chirp span to cover the expected natural linewidth. The 2P resonances are markedly wider at $72\pm5\si{MHz}$, which we attribute to a stronger interaction between the bound electron and local nuclear spins within the 2P wavefunction envelope, as compared to the single donor. Whilst this suggests a dephasing time $<10\si{ns}$ for the current 2P bound electron, using isotopically purified silicon substrate promises to completely suppress nuclear-spin limited decoherence. Thus we expect that dephasing times approaching milliseconds are possible for 2P electron qubits, just as observed already for single donors\cite{muhonen2015quantifying}.}

\begin{figure}[hbtp]
\includegraphics[width=0.48\textwidth]{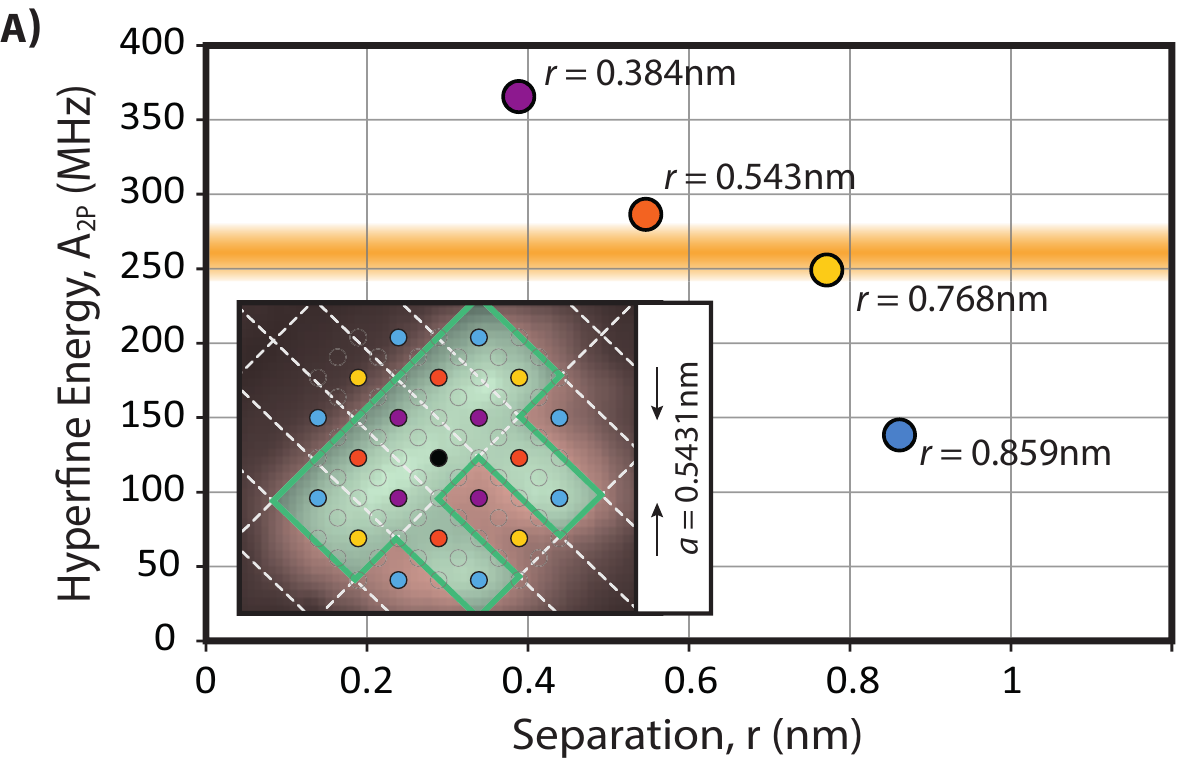}
\caption{\label{fig:three} \textbf{Atomistic tight-binding modelling of the 2P hyperfine energy.}
(A) Simulated hyperfine interaction energy $A_{2P}$, for atomic configurations of a 2P donor molecule with donor separation less than 1nm in the fabrication plane. The dominant uncertainty is due to incomplete knowledge of the Stark shift, reflected in the size of the markers  ($<5\%$). An orange band marks the experimentally observed value of $A_{2P}$ in our device.
The inset shows a schematic of the [001] crystal plane in which the device is fabricated, overlaid on the STM image taken during fabrication (from Fig. 1B)). Atoms in this plane are shown coloured according to their distance from the central black reference site. The marked green zone denotes a potential layout of fully exposed dimers, consistent with the STM image, where it is possible for a $PH_3$ molecule to attach to the surface. Small grey circles represent the location of atoms within the crystal lying in layers above and below the fabrication plane, and white lines indicate the grid of surface dimers.
}
\end{figure}

\subsection{Hyperfine metrology of donor position}

Given the atomic scale of our device, it is possible to model the full electron wavefunctions, accounting for the silicon lattice and bandstructure, donor potentials and the potential profile across the nanostructure.
Hence we compare our measured 2P hyperfine coupling strength, $A_{2P}$, with atomistic tight-binding simulations. From the size of the lithographic patches (Fig 1A,B), we restrict ourselves to consider pairs of lattice sites within a distance of $1\si{nm}$.
Fig. 3A shows the calculated hyperfine energy within a 2P molecule hosting a single electron, where we vary the location and thus the separation between the two P atoms.
The datapoints in Fig. 3A indicate the $A_{2P}$ values for configurations where both donors lie in the [001] crystallographic plane, for 4 different inter-atomic distances $r<1\si{nm}$  (additional out-of-plane donor configurations are discussed for comparison in Supplementary material VI).
\rrevised{The general trend is a reduction in the hyperfine interaction with increasing donor separation, as expected. However the interplay between the tetrahedral symmetry of the silicon crystal lattice around the donor atom, and the cubic symmetry of the 6 conduction band minima in the silicon bulk gives rise to a highly structured 2P wavefunction with deviations from a smooth exponential decay curve that is dependant on the orientation of the donor pair with respect to the crystal lattice\cite{martins2005conduction, wang2016characterizing}.
The dominant uncertainty in the calculated hyperfine coupling A2P is due to the potential Stark shift that could be observed in such a system for the electric field strength and orientation used in the device (see Supplementary Materials VII). The magnitude of the uncertainty ($<5\%$) is reflected by the size of the markers themselves.}

Over the past decade, research on the Si:P system has established that during an anneal at $340\si{\celsius}$, a phosphorus atom incorporates into the surface layer of silicon, forming a P-Si heterodimer with a strong phosphorus-silicon bond \cite{wilson2006thermal}. Consequently, when encapsulated with silicon in a low temperature ($250\si{\celsius}$) epitaxial growth process \cite{goh2004effect}, calculations indicate that the P-Si bond is resilient to segregation or diffusion \cite{bennett2010pathways}. Experimental evidence confirms that the donor atom remains localised \cite{oberbeck2004measurement,usman2016spatial} in the crystal at the lithographically defined site, to an uncertainty on the order of one lattice constant. The orange band represents the hyperfine energy measured experimentally for our 2P molecule, $A_{2P} = 262\pm20\si{MHz}$.
The geometrical layout of the in-plane configurations of donors are displayed in the inset, colour-coded to the hyperfine energy plot. Due to lattice symmetry there are several equivalent sites for the second P atom, at any given distance from the central reference site (coloured black) which represents the location of the first P atom of the molecule.
Considering our experimental value of $A_{2P}$, only the eight geometries coloured orange and yellow are likely representations of the relative configuration of donors in our device. These are consistent with the size of the lithographic patch that was fabricated, as can be seen with reference to the surface dimer reconstruction marked by dashed lines.

\begin{figure}[hbtp]
\includegraphics[width=0.48\textwidth]{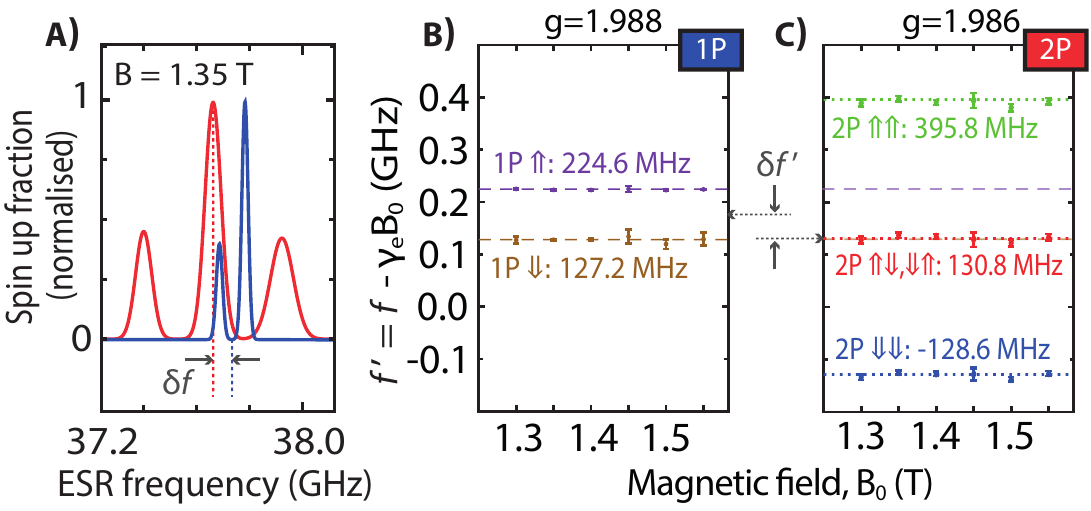}
\caption{\label{fig:four} \textbf{Magnetic field dependence and Stark shift of the electron g-factor.}
(A) Normalised spin resonance spectra for the 1P (blue) and 2P (red) measured at 1.35T plotted together to highlight the relative frequencies.
(B,C) Frequency offsets $f^\prime = f - g_e \mu_B B_0$ as a function of magnetic field $B_0$ for the 2 (3) resonances of the 1P donor (2P molecule) relative to the linear slope of the Zeeman term $g \mu_B B_0 / h$. With a shared y-axis, an offset $\delta f^\prime\sim45\si{MHz}$ due to a Stark shift of the g-factor is evident between the two spectra.
}
\end{figure}

\ssection{G-factors}
An interesting feature observed in our resonance measurements is a relative offset $\delta f$ in the central frequencies, evident when the 1P and 2P spectra are displayed on a single axis as in Fig. 4A. To gain an understanding of this offset, we examine the magnetic field dependence of the resonant frequencies.
Fig. 4B (C) plots the relative offset in resonance frequencies for the 1P electron (2P electron) as a function of magnetic field. The values plotted are obtained from fits to the recorded spectra at each magnetic field setting, with a linear Zeeman term subtracted $f^\prime = f - g_e \mu_B B_0$ for clarity. The electron g-factor used in each figure is obtained by a linear fit to each dataset, producing values of $g_e = 1.988\pm 0.02$ (1P) and $g_e = 1.986\pm 0.02$ (2P). These are both consistent with the bulk g-factor of $1.9985$ for donor-bound electrons \cite{feher1959electron1,stegner2006electrical}, with a $1\%$ uncertainty limited by the current-to-field calibration of our superconducting magnet.

 The offset between the central frequencies of the 1P and 2P spectra $\delta f^\prime\sim45\si{MHz}$ remains constant across the field range. In magnetic field terms this corresponds to $\sim 1.6 \si{mT}$, consistent with the $0.2\%$ difference in g-factor we observe between the 1P and 2P cases. We attribute this  variation to a Stark shift of the g-factor \cite{bradbury2006stark,friesen2005theory}, since the electric field is different in both magnitude and direction for our two quantum dots (see Supplementary material III). \rrevised{Importantly, the difference in peak splitting measured for our 1P and 2P sites $A_{2P} - A_{1P} \sim 165\si{MHz}$, as shown in Fig. 4A, is considerably larger than any variability expected from a Stark shift of the hyperfine coupling or g-factor, for either the single P atom or 2P molecule.}

\section{Discussion}

The range of hyperfine values available for closely separated 2P molecules means that our addressing scheme can be extended to a larger number of qubits. For instance, utilising the four different 2P hyperfine interaction strengths shown in Fig. 3A along with a single donor may produce five individually addressable qubits with unique resonance frequencies, and with an offset greater than inhomogeneous broadening due to nearby nuclear spins, or electric field shifts.
An area requiring further investigation however, is to determine the impact of multiple host $^{31}P$ nuclear spins on coherence times and the overall electron spin dynamics.

We note that hosting single electrons in donor molecules carries a number of additional benefits beyond the intrinsic frequency detuning, including extended $T_1$ relaxation times \cite{hsueh2014spin} and deeper confinement of the two electron charge state. The latter is particularly relevant for implementing singlet-triplet based qubits using donor-bound electrons and to realise SWAP-type two-qubit gates \cite{kalra2014robust,wang2016highly}. The use of larger donor molecules, patterned by STM hydrogen lithography provides additional scope for wavefunction engineering, and permits strong confinement of multi-electron states, as required to achieve Pauli spin blockade for high fidelity state readout \cite{broome2017high}.

%2P stuff to quote in discussion
%gonzalez2014exchange
%klymenko2017electronic

\ssection{Conclusion}
These results represent an important step toward achieving full control over multiple donor spin qubits in silicon. The addressability demonstrated, with frequency detunings an order of magnitude larger than the inhomogeneous linewidth in natural silicon will facilitate selective control over individual qubits with low cross-talk. Combined with isotopically purified $^{28}$Si and NMR control over the nuclear spin states \cite{morton2008solid}, donor molecules provide individual addressability for electron spin qubits, and are attractive for quantum simulation and multi-qubit architectures.

\section{Materials and Methods}

The STM hydrogen lithography was performed in ultrahigh vacuum with an Omicron Variable Temperature instrument. A chemically cleaned Si(001) wafer was annealed at $1100\si{\degree C}$ and passivated in a beam of atomic hydrogen. The hydrogen mask was removed in the required areas by scanning with a tip voltage of around $3-6\si{V}$ and current setpoint of $1-10\si{nA}$. Following lithography, the wafer was dosed with phosphine gas, then heated to $350\si{\degree C}$ to incorporate the P donors \citep{wilson2006thermal} before a $55\si{nm}$ encapsulation layer of epitaxial silicon was grown at a rate of $0.15\si{nm/min}$.
The donor layer was electrically contacted by depositing aluminium onto contact vias formed by reactive ion etching. The contact structures, as well as the microwave antenna were all defined by electron beam lithography using a PMMA mask.

Measurements were performed in a $^3\si{He} / ^4\si{He}$ dilution refrigerator with a base temperature of $50\si{mK}$. A superconducting solenoid magnet provided the external magnetic field. DC voltage signals applied to the gates were generated by Yokogawa 7651 and Stanford Research Systems SIM928 voltage sources. Voltage pulses were generated by a National Instruments USB6363 DAC/ADC device and added to the DC signals with simple resistive voltage dividers. The combined gate control signals were then filtered by two-stage lumped element RC filters inside the dilution fridge with a low-pass cutoff of $150\si{kHz}$, and additional high frequency ($>$GHz) noise was suppressed by distributed 'Eccosorb LS' RF absorber material within the filter enclosure. The microwave signals were supplied to the on-chip antenna from a Keysight E8267D vector signal generator (with phase and pulse modulation signals supplied by a Tektronix 5014C arbitrary waveform generator) via a lossy stainless steel coaxial cable (and additional 1dB attenuator at 4K). The readout signal was collected from the SET by a low noise Femto DLPCA200 transimpedance amplifier and then electrically decoupled and filtered by a Stanford Research Systems SIM910 JFET isolation amplifier and SIM965 Bessel filter before being digitised by the National Instruments USB6363 DAC/ADC.

\rrevised{Our tight-binding method uses the NEMO-3D atomistic solver. The model applies an adjustable cut-off potential at the donor site (a central-cell correction) while elsewhere each donor potential is Coulombic \cite{wang2016characterizing}. This model of the central cell correction has been successful in reproducing the full single donor spectrum, the Stark shifts of the hyperfine coupling [28], as well as the system g-factor, the orbital energies of the bound electrons, and replicating STM based imaging of the donor wavefunction. The Schrodinger and Poisson equations are then self-consistently solved within a 30nm cubic domain to produce the ground-state wavefunction from which the hyperfine interaction strength is computed for each donor.}

%%%%%%%%%%%%%%%%%%%%%%%%%%%%%%%%%%%%%%%%%%%%%%%%%%%%%%%%%%%%%%%%%%%%%
%% The Acknowledgement section
%%%%%%%%%%%%%%%%%%%%%%%%%%%%%%%%%%%%%%%%%%%%%%%%%%%%%%%%%%%%%%%%%%%%%
\section{Acknowledgements}

This research was supported by the Australian Research Council Centre of Excellence for Quantum Computation and Communication Technology (Project No. CE110001027), the U.S. National Security Agency and the U.S. Army Research Office under Contract No. W911NF-13-1-0024.
M.Y.S. acknowledges an Australian Research Council Laureate Fellowship.
This work was performed in part at the NSW Node of the Australian National Fabrication Facility.
S.J.H., M.G.H., E.P. and M.Y.S. designed the experiment; S.J.H., S.K.G., M.B. and J.G.K. fabricated the device; S.J.H. and L.F. carried out spin resonance measurements and analysed the data; C.C.Y., Y.W. and R.R. performed tight-binding simulations. S.J.H. and M.Y.S. wrote the article with input from all authors.
M.Y.S. is Editor-in-Chief of Nature Partner Journals (NPJ) Quantum Information and receives financial benefits from the position. The other authors declare that they have no competing interests.
All data needed to evaluate the conclusions in the paper are present in the paper and/or the Supplementary Materials. Additional data related to this paper may be requested from the authors.
The authors thank Yuling Hsueh, Arne Laucht, Lloyd Hollenburg and Sven Rogge for helpful discussions.

%%%%%%%%%%%%%%%%%%%%%%%%%%%%%%%%%%%%%%%%%%%%%%%%%%%%%%%%%%%%%%%%%%%%%
%% External references
%%%%%%%%%%%%%%%%%%%%%%%%%%%%%%%%%%%%%%%%%%%%%%%%%%%%%%%%%%%%%%%%%%%%%
\bibliography{paper_bib}

\end{document}